\documentclass[a4paper]{article}

\usepackage{icrc2013}
\usepackage[english]{babel}

\newcommand{\be}{\,\begin{equation}}
\newcommand{\ee}{\,\end{equation}}

\title{Self-generated magnetic turbulence and the propagation of galactic cosmic rays}

\authors{
Roberto Aloisio$^{1,2}$ and
Pasquale Blasi$^{1,2}$
}

\afiliations{
$^1$ INAF - Osservatorio Astrofisico Arcetri, Firenze Italy. \\
$^2$ Gran Sasso Science Institute (INFN), L'Aquila Italy.
}

\email{aloisio@arcetri.astro.it, blasi@arcetri.astro.it}

\abstract{Cosmic rays propagating in the Galaxy may excite a streaming instability when their motion is super-alfvenic, thereby producing the conditions for their own diffusion. We present the results of a self-consistent solution of the transport equation where diffusion occurs because of the self-generated turbulence together with a preexisting turbulence injected, for instance, by supernova explosions and cascading to smaller scales. All chemicals are included in our calculations, so that we are able to show the secondary to primary ratios in addition to the spectra of the individual elements. All predictions appear to be in good agreement with observations.}


\begin{document}
\maketitle

\section{Introduction}
After decades of investigation, many aspects of the origin of cosmic rays (CRs) remain unclear. Some of the open problems are related to basic principles, while others are more phenomenological in nature and sometimes it is hard to discriminate between the two. While the motion of CRs in the Galaxy is well described as diffusive, and most observables are qualitatively well described within this approach, the fine structure of the theory is all but well defined. 

A very important piece of the debate on CR diffusion concerns the role of CRs in generating their own scattering centers. This makes the problem of diffusion intrinsically non-linear as was recognized long time ago (e.g. see \cite{Skill,Holm}). The balance of CR induced streaming instability and damping of the self-generated waves leads to conclude that CRs can be confined in the Galaxy by their own turbulence only for energies below a few hundred GeV. Hence, this phenomenon has not received much attention in recent times.

In addition to these problems of principle, related to the basic nature of magnetic turbulence and the interplay between CRs and magnetic fields, there are phenomenological difficulties, raised as a consequence of more and better data. For instance, recent data from the PAMELA and CREAM experiments \cite{Adri,Yoon} provide evidence for a change of slope of the spectra of protons and helium nuclei at rigidity $\sim 200$ GV. The spectrum of protons flattens from $E^{-2.85\pm 0.015}$ (for $E< 230$ GeV) to $E^{-2.67\pm 0.03}$ (for $E> 230$ GeV). Although this feature might result from some poorly known aspects of acceleration in the sources, the most plausible explanation is that something peculiar is happening in terms of propagation of CRs, either in the form of a spatially dependent diffusion \cite{Tomas} or because of a competition between different processes relevant for the evolution of waves \cite{PQ1}. 

Despite these difficulties at all levels, propagation of CRs is usually described by using simple diffusion models implemented in propagation codes such as GALPROP \cite{Strong} and DRAGON \cite{DiBern,Evoli}. In the former, diffusion is isotropic and homogeneous (space independent) and peculiar observational features (for instance changes of slope) are usually modeled by assuming artificial breaks in either the injection spectrum of CRs and/or in the diffusion coefficient. In the latter, some effort has been put in introducing a space dependent diffusion coefficient, but in all other respects it has the same advantages and problematic aspects of the former. Neither approach has currently provided a physical explanation of the recent PAMELA and CREAM data. 

Even independent of these complications, the descriptions of CR transport in the Galaxy have always suffered from what is called the anisotropy problem: the B/C ratio which is sensitive to the propagation model can be equally well explained by one of the following models: 1) {\it Standard Diffusion Model} (SDM), in which the injection spectrum is a power law in rigidity $R$ and the diffusion coefficient (or rather the grammage) is a power law $X(R)\propto \beta R^{-\delta}$ (being $\beta$ the particle velocity) with $\delta\simeq 0.6$ for $R>4$ GV and $\delta=0$ for $R<4$ GV. 2) {\it Reacceleration Model} (RAM), in which CRs suffer second order Fermi reacceleration (typically only important for $R<1-10$ GV) and the injection spectrum is a broken power law $Q(R)\propto R^{-2.4}/\left[1 + (R/2)^{2}\right]^{1/2}$. In both models a break in either the grammage or the injection spectrum are required to fit the data \cite{Ptus,Jone}. In the SDM the strong energy dependence of the diffusion coefficient leads to exceedingly large anisotropy for CR energies $\ge$ TeV \cite{PQ2}. In the RAM the anisotropy problem is mitigated \cite{PQ1} although in both models the detailed shape of the anisotropy amplitude as a function of energy is hardly predictable since it is dramatically dependent upon the position of the few most recent and closest sources \cite{PQ2,Lee,Ptus2}. The injection spectrum required at high energy by the SDM is roughly compatible with the one expected based on diffusive shock acceleration (DSA), while the one required by the RAM is too steep and requires non-standard versions of the acceleration theory, possibly invoking the role of the finite velocity of scattering centers (see \cite{Capri1,Ptus3,Capri2} for a discussion of this issue).

In the present paper we follow up on previous work \cite{PQ1,AloBla} in which it has been proposed that the changes of slope in the spectra of protons observed by PAMELA and CREAM may reflect the interplay between self-generation of turbulence by CRs and turbulent cascade from large scales. The transition from diffusion in the self-generated turbulence to diffusion in pre-existing turbulence naturally reflects in a spectral break for protons from steep (slope $\sim 2.9$) to flatter (slope $\sim 2.65$) at rigidity $R\sim 200$ GV \cite{PQ1}. This break compares well with the protons data as collected by the PAMELA satellite \cite{Adri}, and the high energy slope is in agreement with the one measured by the CREAM balloon experiment \cite{Yoon}. At rigidity $R<10$ GV the spectra show a hardening reflecting the advection of particles with Alfv\'en waves \cite{PQ1}.

A previous attempt at taking into account the self-generated turbulence around the sources of Galactic CRs was made by \cite{Ptus4}, but the effect of pre-existing turbulence was not considered. In \cite{Ptus5} the authors investigate the possibility that a Kraichnan-type cascade may be suppressed at $k\ge10^{-13} cm^{-1}$ as a result of resonant absorption of the waves by CRs. The authors claim that this effect may account for the low energy behaviour of the B/C ratio. 

Since the problem of CR diffusion that we intend to investigate is intrinsically non-linear, it is obvious that many of the oversimplifications discussed above need to be assumed here as well. However we try to make one step towards a physical understanding of the nature of CR diffusion in the Galaxy and of the role of CRs in their own transport. All nuclei from hydrogen to iron (including all stable isotopes) and their spallation and ionization losses are included in the calculations, so that we can compare our results with existing B/C data. We retain the assumption that diffusion may be approximated as homogeneous and isotropic within the propagation region. 

\section{Equation for waves and CR transport}

Following Ref. \cite{PQ1} we consider waves responsible for CR scattering as originated through two processes: 1) turbulent cascade of power injected by supernova remnants (SNRs) at large scales, and 2) self-generated waves produced by CRs through streaming instability. The process of cascading is all but trivial to model, in that it is known that power is channeled in a complex way into parallel and perpendicular wave numbers. It is not the purpose of this paper to explore these aspects in detail. We rather adopt the view that cascading occurs through non-linear Landau damping (NLLD) as modeled in \cite{Zhou}, namely through diffusion in $k$-space with a diffusion coefficient which depends on the wave spectrum $W(k)$ (this reflects the intrinsic non-linearity of the phenomenon). The diffusion coefficient in $k$-space can be written as $D_{kk} = C_{\rm K} v_{\rm A} k^{\alpha_1} W(k)^{\alpha_2}$, where $C_{\rm K}\approx 5.2\times 10^{-2}$ \cite{Ptus6} and $v_{\rm A}=B_{0}/\sqrt{4\pi \rho_{ion}}$ is the Alfv\'en speed in the unperturbed magnetic field (notice that $\rho_{ion}$ is the density of the ionized gas in the Galaxy, not the total gas density). In the present paper we assume a Kolmogorov phenomenology for the cascading turbulence, so that $\alpha_1=7/2$ and $\alpha_2=1/2$, and an unperturbed magnetic field $B_0=1$ $\mu$G. 

The general equation describing wave transport in the absence of wave advection can be written as follows:
\be
\frac{\partial}{\partial k}\left[ D_{kk} \frac{\partial W}{\partial k}\right] + \Gamma_{\rm CR}W = q_{W}(k), 
\label{eq:cascade}
\ee
where $q_{W}(k)$ is the injection term of waves with wavenumber $k$. In the present calculations we assume that waves are only injected on a scale $l_{c}\sim 50-100$ pc, for instance by supernova explosions. This means that $q_{W}(k)\propto \delta (k-1/l_{c})$. The level of pre-existing turbulence is normalized to the total power $\eta_B=\delta B^{2}/B_0^2 = \int dk W(k)$. 

The term $\Gamma_{\rm CR}W$ in Eq. (\ref{eq:cascade}) describes the generation of wave power through CR induced streaming instability, with a growth rate \cite{Skill2}:
\be
\Gamma_{\rm cr}(k)=\frac{16 \pi^{2}}{3} \frac{v_{\rm A}}{k\,W(k) B_{0}^{2}} \sum_{\alpha} \left[ p^{4} v(p) \frac{\partial f_{\alpha}}{\partial z}\right]_{p=Z_{\alpha} e B_{0}/kc} ,
\label{eq:gammacr}
\ee 
where $\alpha$ is the index labeling nuclei of different types. All nuclei, including all stable isotopes for a given value of charge, are included in the calculations. As discussed in much previous literature, this is very important if to obtain a good fit to the spectra and primary to secondary ratios, especially B/C. 

Streaming instability proceeds in a resonant manner, therefore the production of waves with a given wavenumber $k$ is associated with particles with momentum such that their Larmor radius equals $1/k$.  

 \begin{figure*}[!t]
  \centering
  \includegraphics[width=0.45\textwidth]{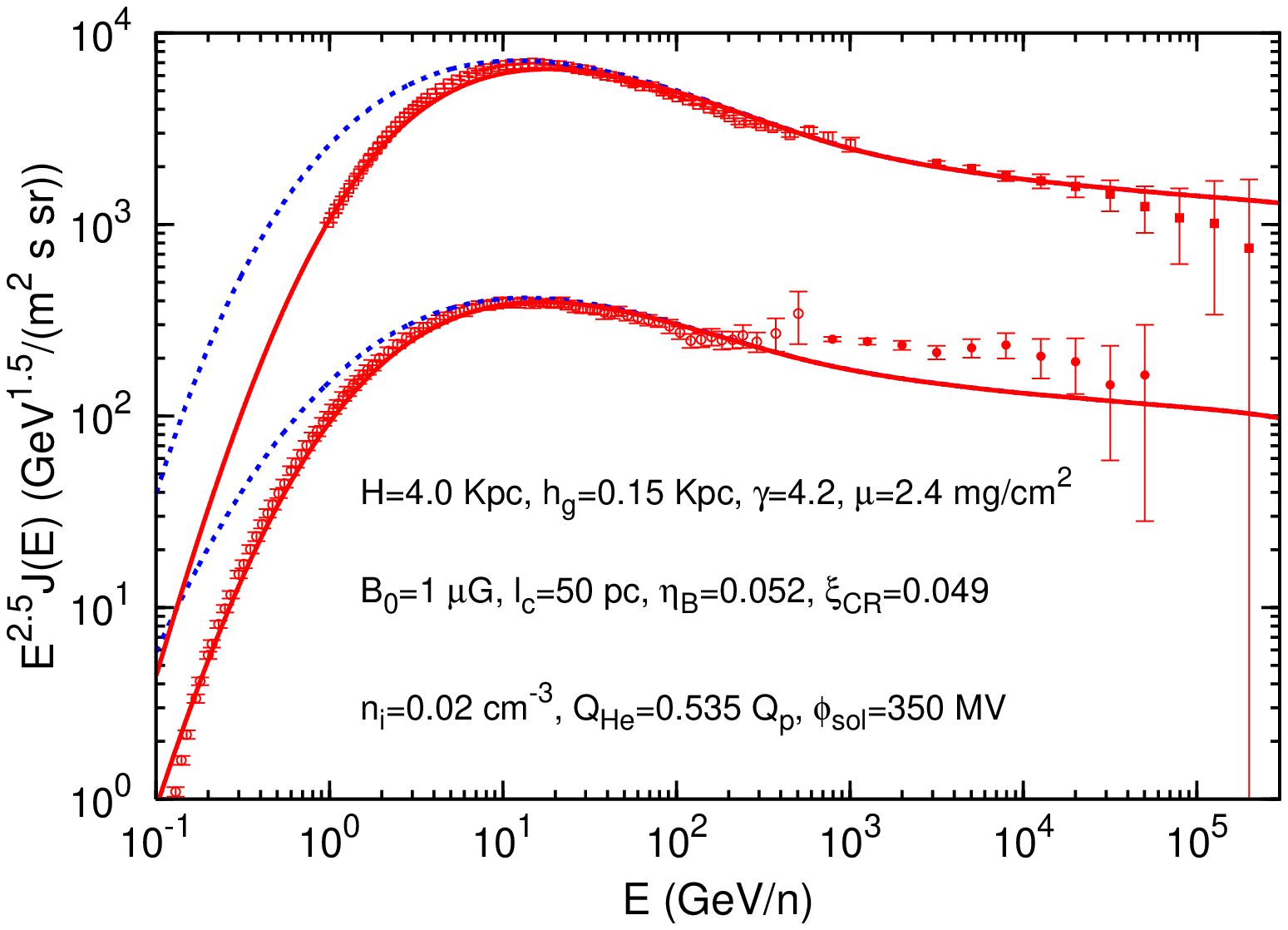}
    \includegraphics[width=0.45\textwidth]{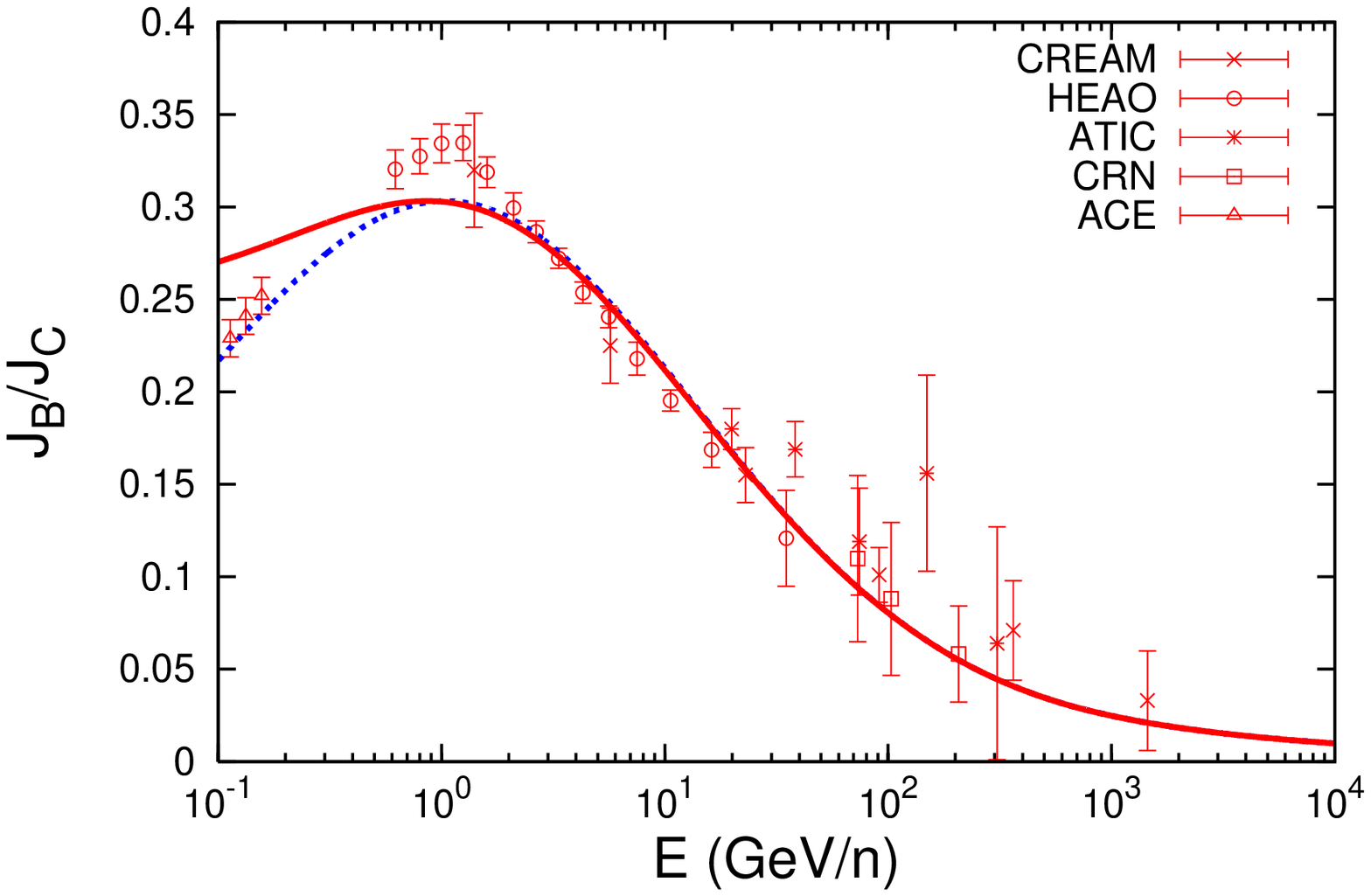}
  \caption{[Left panel] Spectra of protons and helium nuclei for the values of parameters as indicated. The solid (red) lines show the spectra at the Earth, while the dotted (blue) lines show the spectra in the ISM, namely before correction for solar modulation. Data points are from PAMELA \cite{Adri} and CREAM \cite{Yoon,Ahn}. [Right panel] B/C ratio, the solid (red) lines show the spectra at the Earth, while the dotted (blue) lines show the spectra in the ISM. Data points are from CREAM \cite{Ahn}, HEAO \cite{Engel}, ATIC \cite{Panov}, CRN \cite{Swordy} and ACE \cite{George}.}
   \label{fig1} 
 \end{figure*}

A wave with wavenumber $k$ can be either produced or absorbed resonantly by nuclei that satisfy the resonance condition. Hence it is clear that the diffusion coefficient relevant for a given nucleus is the result of the action of all others (see Eq. (\ref{eq:gammacr})). In practice, since protons and helium are the most abundant species, the diffusion coefficients for protons and helium are strongly affected by the presence of both elements. To first approximation, the diffusion coefficient for heavier nuclei is mainly determined by waves produced by protons and helium nuclei. The contribution of all other stable isotopes has a sizable effect, at the level of few percent.

The solution of Eq. (\ref{eq:cascade}) $W(k)$ can be written in an implicit form \cite{AloBla}, being composed by two terms that refer respectively to the pre-existing magnetic turbulence and the CR induced turbulence. The diffusion coefficient relevant for a nucleus $\alpha$ can be written as:
\be
D_{\alpha} (p) = \frac{1}{3} \frac{pc}{Z_\alpha eB_{0}} v(p) \left[ \frac{1}{k\ W(k)} \right]_{k=Z_{\alpha} e B_{0}/pc},
\label{eq:diff}
\ee
where $W(k)$ is the power spectrum of waves solution of Eq. (\ref{eq:cascade}). The non-linearity of the problem is evident here: the diffusion coefficient for each nuclear specie depends on all other nuclei through the wave power $W(k)$, but the spectra are in turn determined by the relevant diffusion coefficient. The problem can be closed, at least in an implicit way, by writing the transport equation for each nucleus. 

The transport equation for nuclei of type $\alpha$ can be written as:
$$
-\frac{\partial}{\partial z} \left[D_{\alpha}(p) \frac{\partial f_{\alpha}}{\partial z}\right] + v_{A} \frac{\partial f_{\alpha}}{\partial z}-
\frac{dv_{A}}{dz}\frac{p}{3}\frac{\partial f_{\alpha}}{\partial p}+\frac{f_{\alpha}}{\tau_{sp,\alpha}}
$$
\be
 + \frac{1}{p^{2}} \frac{\partial}{\partial p}\left[ p^{2} \left(\frac{dp}{dt}\right)_{\alpha,ion} f_{\alpha}\right] =\\ q_{\alpha}(p,z)+\sum_{\alpha'>\alpha} \frac{f_{\alpha'}}{\tau_{sp,\alpha'}},
\label{eq:transport}
\ee
where $\tau_{\alpha,sp}$ is the time scale for spallation of nuclei of type $\alpha$, $q_{\alpha}(p,z)$ is the rate of injection per unit volume of nuclei of type $\alpha$ and $\left(\frac{dp}{dt}\right)_{\alpha,ion}$ is the corresponding rate of ionization losses. Following Ref. \cite{Jone} we introduce a surface gas density in the disc $\mu=2h_{d} n_{d} m$, where $h_{d}$ is the half-thickness of the Galactic disc and $n_{d}$ is the gas density in the disc, in the form of particles with mean mass $m$. The measured value of the surface density is $\mu\approx 2.4 \rm mg/cm^{2}$ \cite{Ferriere}. Since the gas is assumed to be present only in the disc, and the ionization rate is proportional to the gas density, one can write: $\left(\frac{dp}{dt}\right)_{\alpha,ion}=2h_d \delta(z) b_{0,\alpha}(p)$, where $b_{0,\alpha}(p)$ contains the particle physics aspects of the process (see \cite{Strong2} and references therein for a more detailed discussion of this term).  

Integrating Eq. (\ref{eq:transport}) between $z=0^{-}$ and $z=0^{+}$, recalling that $(\partial f/\partial z)_{0^{-}}=-(\partial f/\partial z)_{0^{+}}$, and introducing the function $I_{\alpha}(E)$ (flux of nuclei with kinetic energy per nucleon $E$ for nuclei of type $\alpha$), such that $I_{\alpha}(E)dE=v p^{2} f_{0,\alpha}(p)dp$, it is possible to rewrite Eq. (\ref{eq:transport}) in the weighted slab approximation \cite{Jone}
$$
\frac{I_{\alpha}(E)}{X_{\alpha}(E)} + \frac{d}{dE}\left\{\left[ \left(\frac{dE}{dx}\right)_{ad} +  \left(\frac{dE}{dx}\right)_{ion,\alpha}\right] I_{\alpha}(E)\right\} +
$$
\be
+ \frac{\sigma_{\alpha} I_{\alpha}(E)}{m} = 2 h_d \frac{A_{\alpha} p^{2} q_{0,\alpha}(p)}{\mu v} + \sum_{\alpha'>\alpha} \frac{I_{\alpha}(E)}{m}\sigma_{\alpha'\to\alpha},
\label{eq:slab2}
\ee
being $\sigma_\alpha$ and $\sigma_{\alpha'\to\alpha}$ respectively the total spallation cross section of nucleus $\alpha$ and the spallation cross section of nucleus $\alpha'$ into $\alpha$. The two quantities $X_{\alpha}(E)$ and $\left(\frac{dE}{dx}\right)_{ad}$ are respectively the grammage for nuclei of type $\alpha$ and the rate of adiabatic energy losses due to advection \cite{Jone}. 

The solution of Eq. (\ref{eq:slab2}) can be worked out analytically with the boundary condition $I(E\to\infty)=0$ \cite{AloBla}. 
The injection term in the right hand side of equation (\ref{eq:slab2}) can be written assuming a simple injection model in which all CRs are produced by SNRs with the same power law spectrum:  
$$
2 h_d \frac{A_{\alpha} p^{2} q_{0,\alpha}(p)}{\mu v}=\frac{A_{\alpha} p^2}{\mu v} \frac{\xi_{\alpha}E_{SN}{\cal R}_{SN}}{\pi R_d^2 \Gamma(\gamma) c (m_p c)^4}  \left (\frac{p}{m_p c}\right )^{-\gamma}
$$
where $\xi_{\alpha}$ is the fraction of the total kinetic energy of supernova ($E_{SN}=10^{51}$ erg) channelled into CRs of specie $\alpha$, $R_d=10$ kpc is the radius of the Galactic disk, ${\cal R}_{SN}=1/30$ y$^{-1}$ is the rate of SN explosions and $\Gamma(\gamma)=4\pi\int_{0}^{\infty}  dx x^{2-\gamma}[\sqrt{x^2+1} -1 ]$ comes from the normalization of the CR spectrum. 

 \begin{figure*}[!t]
  \centering
  \includegraphics[width=0.45\textwidth]{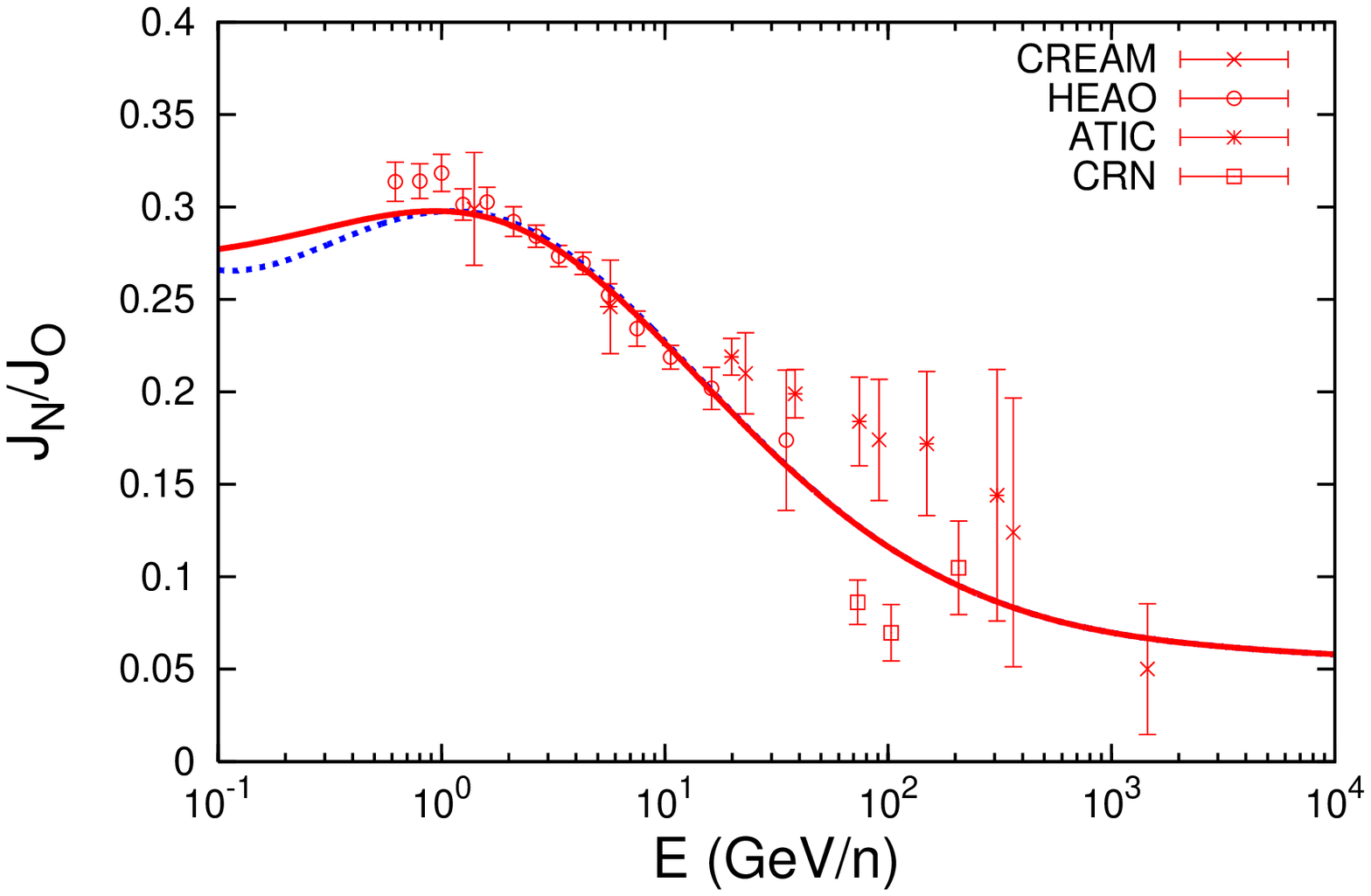}
    \includegraphics[width=0.45\textwidth]{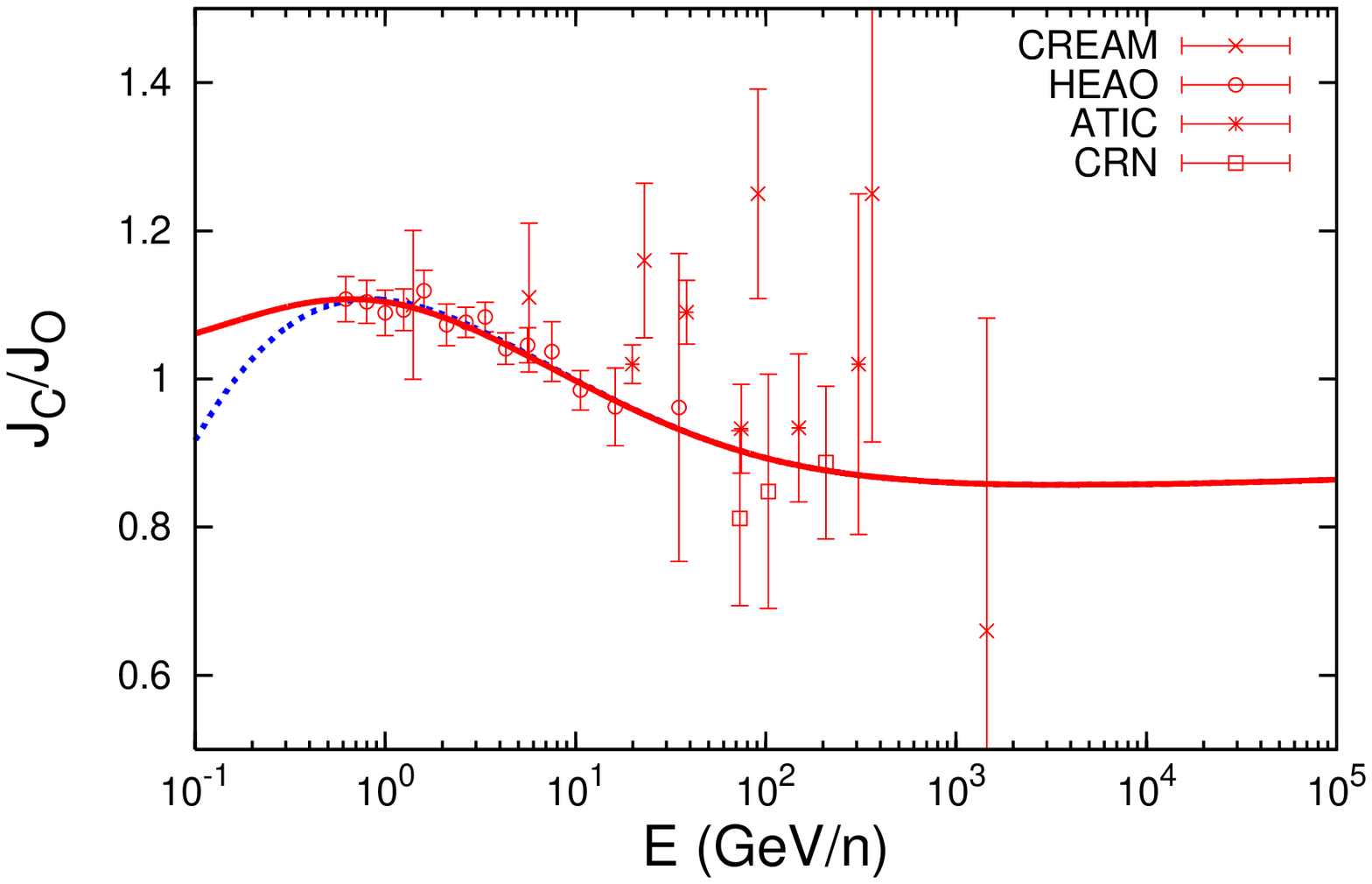}
  \caption{[Left panel] N/O ratio, the solid (red) lines show the spectra at the Earth while the dotted (blue) lines show the spectra in the ISM. [Right panel] C/O ratio. Data points are from CREAM \cite{Ahn}, HEAO \cite{Engel}, ATIC \cite{Panov} and CRN \cite{Swordy}.}
   \label{fig2} 
 \end{figure*}

The equations for the waves (\ref{eq:cascade}) and for CR transport (\ref{eq:slab2}) are solved together in an iterative way, so as to return the spectra of particles and the diffusion coefficient for each nuclear specie and the associated grammage. The procedure is started by choosing guess injection factors for each type of nuclei, and a guess for the diffusion coefficient, which is assumed to coincide with the one predicted by quasi-linear theory in the presence of a background turbulence. The first iteration returns the spectra of each nuclear specie and a spectrum of waves, that can be used now to calculate the diffusion coefficient self-consistently. The procedure is repeated until convergence, which is typically reached in a few steps, and the resulting fluxes and ratios are compared with available data. This allows us to renormalize the injection rates and restart the whole procedure, which is repeated until a satisfactory fit is achieved. Since the fluxes of individual nuclei affect the grammage through the rate of excitation of streaming instability and viceversa the grammage affects the fluxes, the procedure is all but trivial. 

\section{Results}
The calculated spectra of protons and helium nuclei are plotted in the right panel of Fig. \ref{fig1}. The dashed (blue) lines are the spectra in the ISM while the continuous (red) lines represent the spectra after correction for solar modulation, modeled as in \cite{DiBern} with a solar modulation parameter $\Phi=350$ MV. The predicted spectra are compared with the PAMELA \cite{Adri}, CREAM \cite{Yoon,Ahn} and HEAO \cite{Engel} data. The proton spectrum is in excellent agreement with the observed one at all energies and clearly shows a hardening at energies above $\sim 200$ GeV. The helium spectrum is also in good agreement with data up to a few hundred GeV/nucleon. At higher energies the measurements are dominated by CREAM data and the agreement is poorer. Whether this discrepancy is due to a flatter injection of helium nuclei or to a different systematic error between the PAMELA and the CREAM data remains an open question. At low energies we do not see a clear evidence for any harder injection spectrum of helium nuclei compared with protons, as shown by the excellent fit to the PAMELA data. One should recall that a $\sim 20\%$ systematic error in the energy determination in any of these experiments would reflect in a factor $\sim 1.5$ at $\sim $TeV energies in the absolute fluxes as plotted in Fig. \ref{fig1}, due to the multiplication factor $E^{2.5}$.

Using the same choice of parameters labeled in the left panel of Fig. \ref{fig1}, the calculated B/C ratio is compared with data in the right panel of Fig. \ref{fig1} while the N/O and C/O ratios are plotted in Fig. \ref{fig2}. The dashed (blue) lines refer to quantities in the ISM, while the continuous (red) lines show the same quantities after accounting for solar modulation. The steep energy dependence of the B/C ratio observed between 1 and 100 GeV/n is well described by our calculations, as a result of the fact that self-generation is very effective in such energy range.  The quality of the fit to the data is comparable with that obtained by using a standard description of CR propagation (standard diffusion or reacceleration), but here we did not require any artificial breaks in the diffusion coefficient and/or in the injection spectrum. It is also worth stressing that the results illustrated in this paper are not the output of a formal fitting procedure in which the parameter space is scanned systematically and some kind of likelihood method used. Since the calculations are intrinsically non-linear this would be rather time consuming from the computational point of view, therefore we only try to achieve a reasonable fit to the data without actually maximizing a likelihood indicator. 

\section{Conclusions}
Our calculations show that a good fit to the proton and helium spectra (see left panel of Fig. \ref{fig1}) and to the secondary/primary and primary/primary ratios (see right panel of Fig. \ref{fig1} and Fig. \ref{fig2}) can be obtained by properly taking into account the self-generation, the pre-existing turbulence and the advection with the waves. Quite remarkably, the general features of the proton spectrum as measured by PAMELA are very well reproduced. The spectrum is never really a power law, but it may be locally approximated as a power law. The spectrum is rather flat (slope $\le 2$) at energy below 10 GeV/n. It steepens to a slope $\sim 2.8-2.9$ at $10GeV/n\leq E\leq 200GeV/n$ and it gets harder again (slope $\sim 2.6$) at $E>200$ GeV/n. Throughout our calculations the injection spectrum is just a pure power law in momentum, as predicted by the theory of DSA in supernova remnants. The slope of the injected spectrum is $\sim 2.2$ (if the number of injected particles in the range $dp$ of momentum is normalized as $Q(p)dp$). The growth rate of the instability necessary to explain the data requires that CRs be injected in SNRs with a typical CR acceleration efficiency $\sim 5-10\%$. One should appreciate that within this approach we are able to reproduce all observables, most notably the ratios B/C, N/O and C/O, with no need for artificial breaks in injection spectra and/or diffusion coefficient, contrary to what is usually required by standard calculations of CR transport in the Galaxy.

\vspace*{0.5cm}
\footnotesize{

}

\end{document}